\documentstyle[12pt,epsfig]{article}
\begin{document}
\begin{center}
{\bf Analysis of solar neutrino induced $\beta\beta$-processes
for several nuclei}

\bigskip{
\underline {S.V. Semenov}$^{1)}$, Yu.V. Gaponov$^{1)}$, 
F. {\v S}imkovic$^{2)}$, P. Domin$^{2)}$}
\bigskip

{\it 1. Russian Research Center "Kurchatov Institute", Moscow, Russia}\\
{\it 2. Department of Nuclear Physics, Comenius University, 
Bratislava, Slovakia}
\end{center}

\begin{abstract}
We investigate neutrino flux induced $\beta\beta$-transitions
in targets built of $^{112,114,116}$Cd and $^{18}$O 
isotopes. In addition to known $\beta^-\beta^-$ 
channel we consider new $\beta^{-}\beta^{+}$ and $\beta^{-}\beta^{+}\gamma$ 
modes of the neutrino induced $\beta\beta$-process. 
A possibility of detection of the  solar neutrinos 
via the induced $\beta\beta$-transitions of interest is
discussed.We note that the $\beta^-$-part of the solar  neutrino induced 
$\beta^{-}\beta^{+}$-process in $^{18}$O was already
discussed in Ref. \cite{9,10} in connection with a possible 
influence of high energy electron production of this origin 
on the SuperKamiokande results. 

\end{abstract}

The exploitation of stable isotopes in neutrino physics  establishes 
an effective experimental base for investigation of the deficit of
solar neutrinos  and of  different neutrino aspects of various
double beta-processes ($\beta\beta$-processes) including problems of 
the nature and the masses of neutrinos, right-handed 
currents etc. \cite{1,2,3}. 

The theoretical study of the neutrino induced $\beta\beta$-transitions 
\cite{4,5} allows to extend the group of nuclei for experimental
investigation of the $\beta\beta$-processes 
and to develop new  methods for detection
of solar, atmospheric and terrestrial   
(reactor, accelerator and from artificial sources) neutrinos. 
Till now, the attention was paid only to 
processes of spontaneous nuclear $\beta\beta$-decay. 

Recently, it was pointed out that $^{100}$Mo is suitable isotope
for low energy solar neutrino registration \cite{6} 
due to its low reaction threshold of $0.168$~MeV. An analysis
of the associated $\beta\beta$-transitions based on the previous
study for this isotope \cite{5} were carried out in Ref \cite{7}. 
In this contribution we propose another stable isotopes,
in particular $^{112,114,116}$Cd and $^{18}$O, 
for detection of solar neutrinos  via the neutrino induced 
$\beta\beta$-transitions. In addition, new 
$\beta^{-}/EC$ and $\beta^{-}\beta^{+}$ channels \cite{7}  of 
the induced $\beta\beta$-process are examined.

The neutrino induced $\beta\beta$ transition
\begin{eqnarray}
\nu_{e} + (A,Z) \rightarrow (A,Z+2) + 2e^{-} + \bar{\nu}_{e},
\label{eq:1}
\end{eqnarray}
is a second order process in the weak interaction, which is
allowed within the Standard model. The amplitude of this process 
exhibits a resonant behavior and depends strongly on widths
the intermediate nuclear states. 

The total cross section $\sigma (\varepsilon_{\nu})$
of the neutrino induced $\beta\beta$-process can be written as~\cite{4,7}:
\begin{equation}
\sigma (\varepsilon_{\nu}) = 
\sum\limits_{m=0}^{m^{RS}} \sigma_{\beta}^{(m)}(\varepsilon_{\nu })
\frac{\gamma_{mf} }{\gamma _{m} },
\end{equation}
where $m^{RS}$ denotes the highest lying 
real excited state of the intermediate 
nucleus, which is allowed by the energy conservation.
$\sigma_{\beta }^{(m)} (\varepsilon_{\nu })$ is a
cross section of the neutrino induced single $\beta$-process 
associated with transition from the ground state of the initial nucleus 
to the $m$-th excited state of the intermediate nucleus.
The total and the partial (related with $\beta^+$, $\beta^-$ and 
EC decay channels) widths  of this state are denoted as
$\gamma_{m}$ and $\gamma_{mf}$, respectively. It is worthwhile to notice
that in the case of the excited states of the intermediate nucleus
the ratio $\gamma_{mf}/\gamma_{m}$ is small due to large 
corresponding electromagnetic widths. It means that
the induced $\beta\beta$-process through excited intermediate 
nuclear states without additional emission of $\gamma$-ray 
is strongly suppressed. This fact points out the importance
of the $\beta\beta$-process with $\gamma$-ray emission 
associated with electromagnetic transitions  
from the excited to the ground state of the intermediate 
nucleus  and subsequent single $\beta$-transition to the 
final state. 

The cross section $\sigma_{\beta}^{(m)} (\varepsilon_{\nu } )$
can be expressed with help of the $\log ft_{\beta^{+} ,EC}$
value as follows:
\begin{eqnarray}
\sigma_{\beta }^{(m)}(\varepsilon_{\nu}) 
&=& \frac{2\ln 2 (2J_{m} +1) \pi^{2}}{m_{e}^{3} 10^{\log ft_{\beta {+} ,EC}} }
\pi_{r}^{(m)} \varepsilon_{r}^{(m)} F(Z,\varepsilon_{r}^{(m)}) \nonumber \\
&= & \frac{0.2625(2J_{m} +1)}
{10^{\log ft_{\beta^{+} ,EC}} } \pi_{r}^{(m)} 
\varepsilon_{r}^{(m)} F(Z,\epsilon_{r}^{(m)}) \cdot 10^{-40} 
\mathrm{cm}^{2}.
\end{eqnarray}
Here, $J_{\mathit{m}}$ is the angular momentum of the 
$m$-th excited state of the intermediate nucleus.
$\log ft_{\beta^{+} ,EC}$ is related with
the $\beta$-transition from the $m$-th state 
of the intermediate nucleus to the initial nucleus. 
$\pi_{r}^{(m)}$ and $\varepsilon_{r}^{(m)}$
are the momentum and the energy of outgoing electron 
in units of the mass of electron $m_e$, respectively.
$\varepsilon_{r}^{(m)} =\varepsilon_{\nu} -\varepsilon_{m} +\varepsilon_{i}$, 
where $\varepsilon_i$ and $\varepsilon_m$ are energies 
of the initial and the intermediate nuclear states.  
$F(Z, \varepsilon_{r}^{(m)})$ is the Coulomb correction function. 
The neutrino threshold energy for the induced $\beta\beta$-process
is given by:  $\varepsilon_{\nu, thr} = \varepsilon_m - \varepsilon_i + 1$.

The production rate for solar neutrino events in $10$~tones of $^{100}$Mo 
per day were calculated in Ref. \cite{7}. The expected experimental signal 
consists of two emitted electrons with the time delay of 15.8 seconds. 
The subtraction of the two-neutrino double beta decay
($2\nu\beta\beta$-decay) background
seems to be a serious problem for this type of detector.
However, they could be eliminated by the coincidence measurements as in 
the $2\nu\beta\beta$-decay two  electrons are emitted simultaneously.

In this work we present solar neutrino absorption and production rates 
for $^{112,114,116}$Cd and $^{18}$O isotopes. 
The absorption rate for a given  component $s$ of  the solar neutrino 
flux takes the form \cite{8}
\begin{equation}
R_{\nu } =\int \sigma^{(m)}(\varepsilon_{\nu}) 
\rho_{s}(\varepsilon_{\nu}) d\varepsilon_{\nu}.
\end{equation}
Here, $\rho_s(\varepsilon_{\nu})$ is 
the energy distribution of solar neutrinos of the origin $s$.

\begin{table}[t]
\caption{Solar neutrino absorption rates $R_{\nu}$ in SNU 
in $^{112}$Cd, $^{114}$Cd and $^{116}$Cd 
and production rates $I$ in $10$~tones of these isotopes per day
for induced $\beta\beta$-transitions. 
$E_{\nu, thr}$ and $\tau$ denote solar neutrino threshold 
energy for induced $\beta\beta$-process for a given isotope
and the expected time delay between the emission of two
electrons in this process, respectively. $m$ and $s$ stand
for minutes and seconds, respectively.
\label{tab:1}}
\begin{tabular}{|l|l|l|l|ll|l|}
\hline
Nucleus& 
$E_{\nu, thr}$&
$\log ft_{EC}$& 
$\tau$&
\multicolumn{2}{l|} {$R_{\nu}$}& 
$I$\\
&[MeV]&&&\multicolumn{2}{l|}{[SNU]}&[$day^{-1}$ $(10~ton)^{-1}$]\\
\hline
$^{112}$Cd& $2.578$& $4.7$& $14.4$~m& $^{8}$B: &$3.49$& 
$0.016\left\{
\begin{array}{l}
0.007~\beta^{-} \beta^{-}\\ 
0.004~\beta^{-} \beta^{+}\\
0.005~\beta^{-} / EC
\end{array}\right.$
\\
\hline
$^{114}$Cd& 
$1.444$& 
$4.9$& 
$71.9$~s& 
$^{8}\mathrm{B}$:& $3.04$& 
$0.014$\\
&&&&$^{15}\mathrm{O}$:&$0.46$&$0.002$ \\ 
\hline
$^{116}$Cd& $0.465~(1^+_{g.s.})$& 
$4.39~(1_{g.s.}^{+})$&
$14.1$~s& 
$^{7}{\mathrm{Be}(2)}$:& $237.11$&
$1.06$\\ 
&$1.465~(1^+_1)$&$4.98~(1^+_1)$&&
$^{8}{\mathrm{B}}$:& $1_{g.s.}^{+}~12.63$&
$0.057$\\
&$2.665~(1^+_2)$&$4.83~(1^+_2)$&&
&$1_{1}^{+}~2.49$&
$0.011$\\
&&&&
&$1_{2}^{+}~2.49$&
$0.011$\\
&&&&
$^{13}{\mathrm{N}}$:& $18.39$&
$0.083$\\
&&&&
$^{15}{\mathrm{O}}$:& $37.08$&
$0.166$\\
\hline
\end{tabular}
\end{table} 

In Table \ref{tab:1} we present results for cadmium nuclei. 
The production rates $I$ were obtained for detector 
consisting of $10$~tones of a given cadmium isotope. 
We considered only those components of the solar neutrino spectrum, 
which give significant contributions to solar neutrino absorption
rates. In the case of $^{112,114}$Cd only transitions through the
ground state of the intermediate nucleus were taken into account
as contributions from transitions through excited states 
are negligible. There is a different situation in the case
of $^{116}$Cd. For a correct evaluation  of the $^{8}$B solar 
neutrino induced $\beta\beta$ process in this isotope
it is important to consider also transitions through 
the $1_{1}^{+}$ and $1_{2}^{+}$ excited states of 
$^{116}$In, which are accompanied with $\gamma$-ray emission.
We note that $\log ft_{EC}$ values associated with $\beta$-transition 
from the excited states of $^{116}$In were calculated from Gamow-Teller
strengths of $^{116}$In measured by
${^{116}\mathrm{Cd}}({^{3}\mathrm{He}}){^{116}\mathrm{In}}$ reaction in
Ref.~\cite{aki97}.
From the Table \ref{tab:1} we can conclude  that
$^{116}$Cd is a good candidate  
for detection of Be(2) solar neutrino flux. 

The solar  neutrino induced $\beta\beta$-transition
introduced in (\ref{eq:1})  is the dominant mode  
in the case of $^{114}$Cd and $^{116}$Cd. It is
worthwhile to notice that there are two additional
channels of this process within the Standard model,
namely $\beta^{-}\beta^{+}$  and $\beta^{-}/EC$ modes \cite{7}:
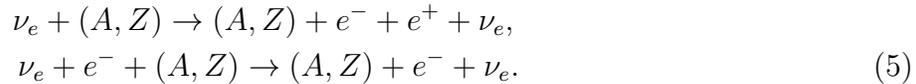
\begin{eqnarray}
\nu_{e} + (A,Z) \rightarrow (A,Z)+e^{-} + e^{+} + \nu_{e}, \nonumber\\
\nu_{e} + e^{-} + (A,Z) \rightarrow (A,Z) + e^{-} +\nu_{e}.  
\end{eqnarray}
The branching ratios for $\beta^{-}\beta^{+}$  and 
$\beta^{-}/EC$ channels of the solar neutrino induced 
$\beta\beta$-process in $^{114}$Cd are of the order of 
$1.9$ and $0.004$ percents, respectively. 

In the case of $^{100}$Mo the contribution 
from the $\beta^{-}\beta^{+}$-channel
to the full cross-section of the $\beta\beta$-process is
strongly suppressed. It is due to the fact that 
this transition is realized  through excited 
states $m$ of the intermediate nucleus $^{100}$Tc for
which the ratio $\gamma_{m\beta^{+}}/\gamma_{m}$
is negligible small due to large 
associated electromagnetic widths $\gamma_{\mathit{m}}$ 
\cite{7}. In the case of $^{112}$Cd there is  
a different scenario as the ground state of $^{112}$In 
is unstable in respect to the $\beta^-$- and $\beta^+$-decays 
and electron capture
($\beta^{-}$~($44\%)$, $EC$~$(22\%)$ and $\beta^{+}$~$(34\%)$).
Thus,  the $^{8}$B solar neutrino induced $\beta^{-}\beta^{+}$  
transition in $^{112}$Cd is allowed. However, the counting rate
for emission of positrons is low due to small neutrino 
absorption cross section for this isotope. 

It is worth mentioning that the solar  neutrino registration 
with help of the emitted positrons in the induced 
$\beta^{-}\beta^{+}$-process is favored.
The annihilation of positron with atomic
electron leads to a signal of two simultaneously emitted
$\gamma$-ray with significant time delay in respect to emission of 
the first electron. This fact allows considerably to reduce 
background events. However, it is desirable  that
in the detection of solar neutrinos via induced 
$\beta^{-}\beta^{+}$-transition a target with 
large values of inverse beta-decay cross section and branching 
ratio for $\beta^{+}$-decay from the intermediate nucleus
is chosen. We find out  $^{18}$O to be a good candidate
for this purpose. The open decay channels of the
ground state of the neighbor nucleus $^{18}$F are $\beta^+$-decay
and electron capture with branching rates of the order of 
$(97\%)$ and $(3\%)$, respectively. A rather high abundance $(0.2\%)$ 
and effective technology of its production allow to get this 
isotope in sufficient amount for solar neutrino experiments. 
The threshold energy for absorption of solar neutrinos  
in  this isotope 
is $E_{\nu, thr}=1.655$~MeV, what is significant less in comparison
with the threshold energy of $6.5$~MeV for detection of solar neutrinos
in the  SuperKamiokande experiment. 

The above arguments together with large value of the cross section 
for neutrino induced $\beta^{-}\beta^{+}$-process
and expected suppressed background mostly  due to the absence of 
$2\nu\beta\beta$-decay events suggest $^{18}$O  to be a 
perspective tool for investigation of $^{8}$B solar neutrino 
flux in view of the measured discrepancy between Standard solar
model  prediction 
and SuperKamiokande measurement~\cite{8}. The relevant processes
are given by
\begin{equation}
\nu_{e} + ^{18}\mathrm{O} \rightarrow ^{18}\mathrm{O} +e^{+} + e^{-} + \nu_{e}, 
\label{eq:3}
\end{equation}
\begin{equation}
\nu_{e} + ^{18}O \rightarrow ^{18}O + e^{+} + e^{-} + \nu_{e} + 
\gamma~(1.04~\mathrm{MeV}). \label{eq:4}
\end{equation}
Here, we assume that in reaction (\ref{eq:4})
the emitted $\gamma$-ray originates from the electromagnetic 
disintegration of excited states of $^{18}$F, i.e., its energy is fixed
by the energy difference of corresponding nuclear levels.
This electromagnetic radiation
is expecting to be a useful signal in detecting of solar neutrinos.

\begin{table}[t]
\caption{Solar neutrino absorption rate $R_{\nu}$ in SNU 
and production rates $I$ in $1000$~tones of $^{18}$O 
per year associated with the induced
$\beta^{-}\beta^{+}$-processes in Eqs. (6) and (7).
\label{tab:2}}
\begin{tabular}{|l|l|l|l|l|l|}
\hline
Reaction& 
$E_{\nu, thr}$~[MeV]& 
$\log ft$& 
$\tau$& 
$R$~[SNU]& 
$I$~[$(1000~ton)^{-1}$ $year^{-1}$])\\
\hline
(\ref{eq:3})& 
$1.655$& 
$3.57$& 
$109.77$~m& 
$20.06$& 
$20535$\\
\hline
(\ref{eq:4})& 
$2.695$& 
$3.47$& 
$109.77$~m& 
$6.11$& 
$6259$\\
\hline
\end{tabular}
\end{table}

The dominant contributions to the cross section of the  neutrino 
induced $\beta^{-}\beta^{+}$-process  in $^{18}$O comes from
the transitions through $1_{g.s.}^{+}$ ground state
and $0^{+}$ excited states of $^{18}$F. 
In that contribution the transition through $1^+_1$ excited state of
$^{18}$F is not taken into account as its contribution to the absorption 
rate is expected to be small.
We note that the $\beta^-$-part of the solar  neutrino induced 
$\beta^{-}\beta^{+}$-process in $^{18}$O was already
discussed in Ref. \cite{9,10} in connection with a possible 
influence of high energy electron production of this origin 
on the SuperKamiokande results. 

The calculated production rates for solar neutrino events 
in $1000$~tones of $^{18}$O per year associated with reactions 
(\ref{eq:3}) and (\ref{eq:4}) are listed in Table \ref{tab:2}.
By glancing the results we conclude that the neutrino detector, 
which uses the advantage of registration of neutrinos
with help of the neutrino induced $\beta^{-}\beta^{+}$-process 
in $^{18}$O, could be employed for investigation of different
problems of the weak interaction and the nuclear physics,
in particular of the solar neutrino deficit and of 
determining of  neutrino component of the reactor antineutrino flux. 
The experimental signal consists of an emitted electron 
followed with two time delayed  $\gamma$-quanta due
to positron-electron annihilation and in the case of reaction
(\ref{eq:4}) a $\gamma$ ray with energy $1.04~\mathrm{MeV}$
coming from the electromagnetic 
de-excitation of the intermediate nucleus $^{18}$F. 
From Table \ref{tab:2} it follows that the intensity of the
reaction (\ref{eq:4}) is about 10 times smaller in comparison
with that for reaction (\ref{eq:3}).

In summary, a possibility of the  registration of solar neutrinos
via different neutrino induced $\beta\beta$-processes  were addressed.
The calculations were performed for several nuclear systems of
interest. The role of the neutrino induced $\beta\beta$-processes
accompanied with emission of $\gamma$-ray was discussed. We have
found that $^{18}$O is a good candidate for study of the solar
neutrino deficit with help of the neutrino induced 
$\beta^{-}\beta^{+}$-process.

This work  has been supported by the Grant Agency of the Czech 
Republic under contract No. 202/02/0157.

\end{document}